\documentstyle[prc,aps,twocolumn,epsfig]{revtex}
\def\Journal#1#2#3#4{{#1} {\bf #2}, #3 (#4)}  
  
\def\NCA{{\em Nuovo Cimento} A}

\def\EPJA{{\em Eur. Phys. J.} A}  
\def\PLB{{\em Phys. Lett.}  B}  
\def\PRL{\em Phys. Rev. Lett.}  
\def\PRD{{\em Phys. Rev.} D}  
\def\PRC{{\em Phys. Rev.} C}

\begin{document}
\draft

\title{   A Search for Neutral Baryon Resonances Below Pion Threshold   
 }  
\author{X.~Jiang \cite{current_jiang},  R.~Gilman, R.~Ransome}
\address{
 { \it Department of Physics and Astronomy,
Rutgers University, Piscataway, New Jersey 08854.}
}
\author{P.~Markowitz}
\address{
{\it Department of Physics, Florida International University, Miami, Florida 33199.}
}
\author{T.-H.~Chang}
\address{
 {\it Department of Physics, University of Illinois at Urbana-Champaign, Champaign, Illinois 61820.}
}
\author{C.-C.~Chang}
\address{
 {\it Department of Physics, University of Maryland, College Park, Maryland 20742.}
}
\author{G.~A.~Peterson}   
\address{
{\it Department of Physics, University of Massachusetts at Amherst, Amherst, Massachusetts 01003.}  
}
\author{D.~W.~Higinbotham, M.~K.~Jones, N.~Liyanage \cite{current_nilanga}, J.~Mitchell, B.~Wojtsekhowski}
\address{
 {\it Thomas Jefferson National Accelerator Facility, Newport News, Virginia 23606.}
}

\date{August  12, 2002}

\maketitle

\begin{abstract}
The reaction $p(e,e^{\prime}\pi^+)X^0$ was studied with two
 high resolution magnetic spectrometers to search for narrow baryon resonances.
A missing mass resolution of 2.0 MeV was achieved.
 A search for structures in the mass region of $0.97<M_{X^0} <1.06$ GeV yielded no
significant signal. 
The yield ratio of $p(e,e^{\prime}\pi^+)X^0/p(e,e^{\prime}\pi^+)n$ 
was determined to be $(-0.35 \pm 0.35) \times 10^{-3}$ at 1.004 GeV and
$(0.34 \pm 0.42) \times 10^{-3}$ at 1.044 GeV.
\end{abstract}

\narrowtext

Since the discovery of the $\Delta(1232)$ resonance in the 1950's,  
many baryon resonances have been discovered with $m>m_{\Delta}$.   
These baryon states are interpreted as net three-quark 
color-singlet objects with angular and radial excitations~\cite{feynman}. 
The quark model explains the mass difference  
 of $m_\Delta -m_N$ by quark spin-spin interactions~\cite{close}. 
Since all the low-lying quark-model states are accounted for, a 
baryon bound state with a mass between the
nucleon and the $\Delta$ should not exist in the present theoretical 
framework.  Indeed, there was no evidence for such a resonance prior to 1997, 
and several searches~\cite{sram,naka} for charge-two  
resonances yielded null results.   
In 1997, however, possible evidence of neutral baryon states  
at 1.004, 1.044, and 1.094 GeV was reported in the $pp \rightarrow p \pi^+ X^0$ reaction~\cite{tati}.  
The two lower mass states are below pion threshold, and the only allowed
decay channels are the radiative ones, which implies that their  natural widths  
are of the order of a few keV,  much narrower than the experimental resolution of a few MeV.
The authors suggested a possible explanation of these resonance states 
in terms of interacting colored quark clusters.

 These experimental results are most astounding when 
 one considers the countless experiments carried out with many 
different probes over more than 50-years in which 
the claimed states were never observed.
L'vov and Workman~\cite{workman}  argued that the reported 
structures are ``completely excluded''  
by the fact that no such structure was reported in the existing 
real Compton scattering data~\cite{sal}. 
Furthermore,  the existence of these states appears to be ruled out by their  
effects on the predicted composition of a neutron star which lead to a reduced maximal 
mass inconsistent with the observational limit~\cite{kolo}. 
As a counter argument, Kobushkin \cite{kobu} 
suggested that the claimed  states could  be members of a total anti-symmetric 
representation of a spin-flavor group such that the one-photon excitation or decay 
channels are forbidden and only the $2 \gamma N$ channels are allowed.
While there is  no room for these new exotic baryon states 
within  the many theoretical constituent quark models \cite{capstick}, a colorless
 Diquark Cluster Model \cite{konno1} mass formula closely  reproduced the observed masses. 
Recently, a model based on the excitation of 
quark condensates \cite{walcher} was suggested which interpreted
 the resonances as multiple production of
a ``genuine'' Goldstone Boson with a mass of 20 MeV.  

\begin{figure}[tbhp] 
\centerline{
{\epsfig{figure=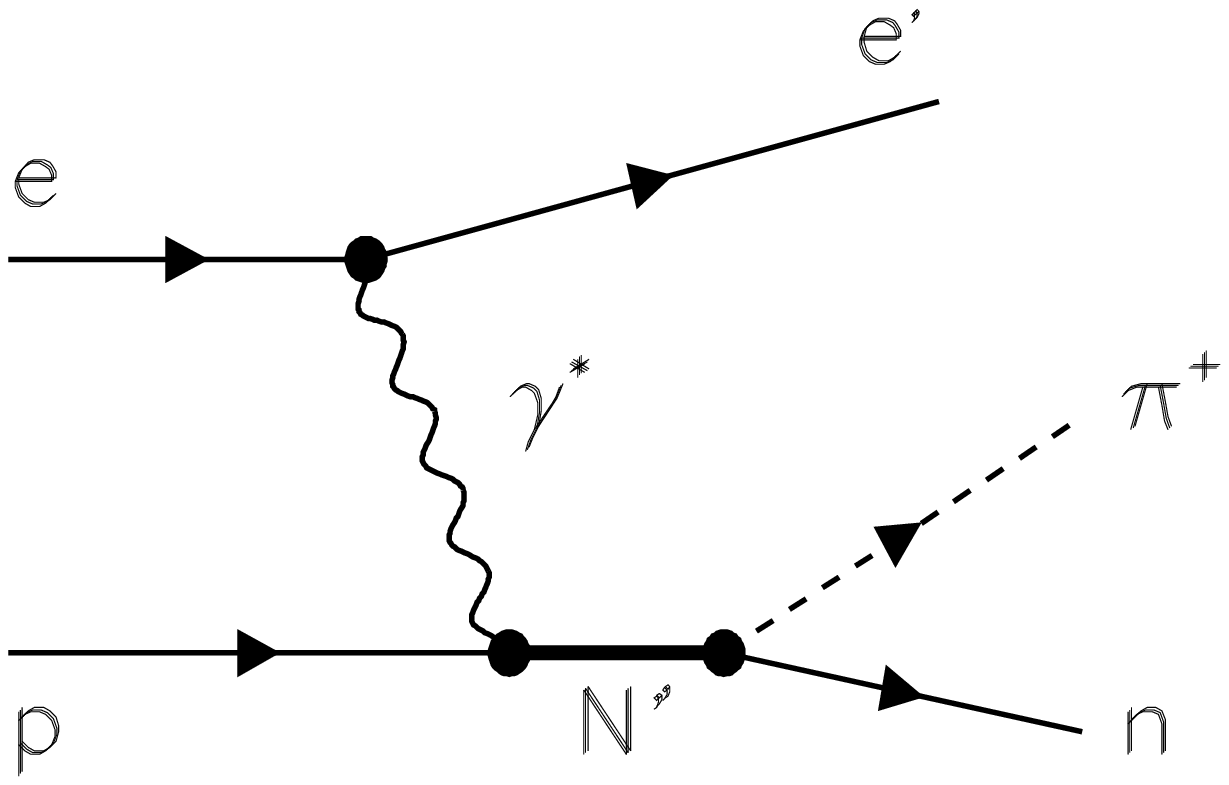,angle=0,height=28mm}} {\epsfig{figure=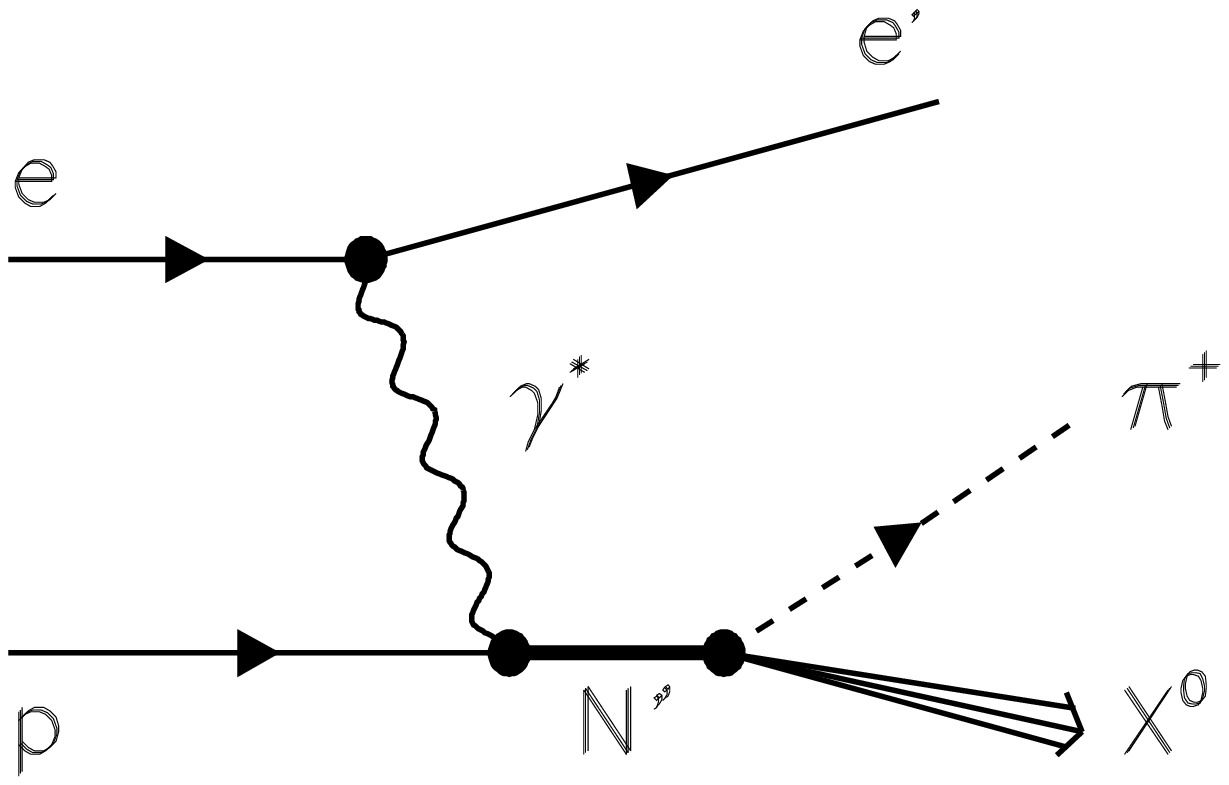,angle=0,height=28mm}} 
}
\vspace{0.5cm}
\caption{\label{fig:expp} The Feynman diagrams of $p(e,e^{\prime} \pi^+)n$ and  $p(e,e^{\prime} \pi^+)X^0$  
reaction.}  
\end{figure}

The existence of baryon resonances below pion threshold, if established experimentally,
could profoundly change our understanding of  quark-quark interactions and strongly
suggest new degrees of freedom in the quark model.  
However, the states claimed in Ref.~\cite{tati}
were of limited statistics amid a rather significant background.  
The signals of the  $pp \rightarrow p \pi^+ X^0$ peaks 
were of the order of $\sim 10^{-3}$ compared to that of the $pp \rightarrow p \pi^+ n$ peak.
Given the potential impact of these states, experimental verification
in different reaction channels is highly desired.  Recently, single baryon states 
of 0.966, 0.987, and 1.003 GeV were reported  in the missing mass spectra of the 
$pd \rightarrow ppX^0$  reaction~\cite{filkov}, but a similar search 
in the same reaction channel has reported no resonance structure \cite{tamii}. 
This paper
reports the first dedicated search in the $p(e,e^{\prime} \pi^+)X^0$ channel
in the mass region of $0.97<M_{X^0}<1.06$ GeV.
As illustrated in Fig.~\ref{fig:expp}, the $X^0$ states 
would be a product of strong interactions through an intermediate state $N^{\prime}$
of the nucleon, $\Delta$, or  N$^\ast$ resonances. 

The ratio of the coupling at the $\pi N^{\prime} X$ vertex to that at the
$\pi N^{\prime} n$  vertex can be determined through the cross section ratio:
\begin{equation}
 \left( {g_{\pi N^{\prime} X} \over g_{\pi N^{\prime} n}} \right)^2={\sigma_{p(e,e^{\prime} \pi^+)X^0} \over \sigma_{p(e,e^{\prime} \pi^+)n}} \equiv K_s.
\end{equation}
According to B. Tatischeff {\it et al.}, the suppression factor 
$K_S$ is expected to be in the order of $\sim 10^{-3}$.
Therefore, a high-resolution, high-statistic and low background 
$(e,e^{\prime} \pi^+)$ measurement could be used
to reveal the resonance states as abnormal structures above the radiative tail
of the $p(e,e^{\prime} \pi^+)n$ reaction, as first suggested by Azimov\cite{azimov}. 

The measurement was conducted at the Thomas Jefferson
National Accelerator Facility's experimental Hall A, taking advantage of
the high resolution spectrometer pair and the high quality CW beam of the 
CEBAF accelerator. The data were collected during a 12 hour period.
An electron beam of  energy 1.722 GeV and average current 33 $\mu$A 
was scattered on a liquid hydrogen target. The target cell was an aluminum can 
15 cm in length and 6.35 cm in diameter, and was oriented with its axis along the beam direction.
Two magnetic spectrometers were used in coincidence. One spectrometer was 
set to a central momentum of 1.040 GeV/c to detect the scattered electrons 
at 19$^\circ$ on beam right,
the other spectrometer was set to 41.6$^\circ$ on beam left to detect $\pi^+$ particles
first at 0.621 GeV/c (Kinematics-A) corresponding to the missing neutron peak  
 as a calibration, then at 0.543 GeV/c (Kinematics-B) for the resonance search.
The four-momentum transfer square is 0.2 GeV$^2$, and  the 
invariant mass of the $\pi^+ X^0$ system is $W\approx1.44$ GeV.  

\begin{figure}[ht]
\centerline{\psfig{figure=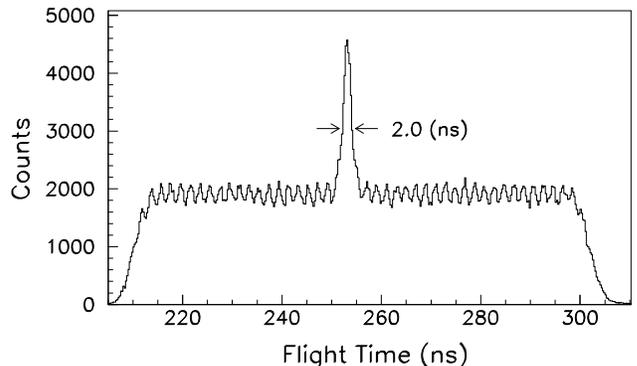,height=50mm}}
\caption{The time-of-flight spectrum in Kinematics-B. The 2.0 ns 
beam structure is clearly visible in the accidental events. Particle ID cuts and vertex cuts have been applied.}
\label{fig:ctof}
\end{figure}

A combination of a threshold gas 
Cherenkov counter and a lead-glass calorimeter array 
at the focal plane provided clean $e/\pi^-$ separation in the electron arm.
In the hadron arm, the  $\pi^+/p$ separation 
was achieved by an Aerogel Cherenkov detector combined with the particle's
 velocity and  energy loss measured by the  trigger scintillators. 
The resolution of the reconstructed two-arm vertex was $\sigma_z =0.6$ cm along the beam direction,
and was mainly limited by the multiple scattering through the window material at the target 
chamber and the spectrometer entrance.
The path length corrected coincidence time had a resolution of $2.0$ ns (FWHM) which is dominated
by the rise time of the photomultiplier tubes attached to the scintillators. 
After particle ID cuts and a two-arm vertex cut of $|\Delta z| < 1.0$ cm, 
the ratio of real-to-random coincidences was about 1:1 in Kinematics-B  
as shown in Fig.~\ref{fig:ctof}.  

Yields were corrected for event-reconstruction efficiencies 
($\sim90\%$) and data acquisition dead times ($\sim10\%$). The effect of pion decay
was accounted for by 
 weighting each event with a survival factor
which was calculated based on the pion's momentum and path length ($\sim$23.5 m).
A window of 3.4 ns centered around the timing peak was used to select the 
$(e,e^{\prime} \pi^+)$ events, and accidental events were sampled from  
a 34 ns time window on each side of the timing peak.  
No noticeable fine-structure was observed in any spectrum of accidental events.
The phase space volumes were calculated through a Monte Carlo simulation which started by 
sampling a missing mass range uniformly. 
Spectrometer models, reconstruction resolutions and target 
material effects were considered in the simulation.
The charge-normalized $(e,e^{\prime} \pi^+)$ yields 
were obtained by subtracting the accidental events
from the coincident events and dividing the result by the phase space volume.

\begin{figure}[hptb] 
\centerline{
\epsfig{figure=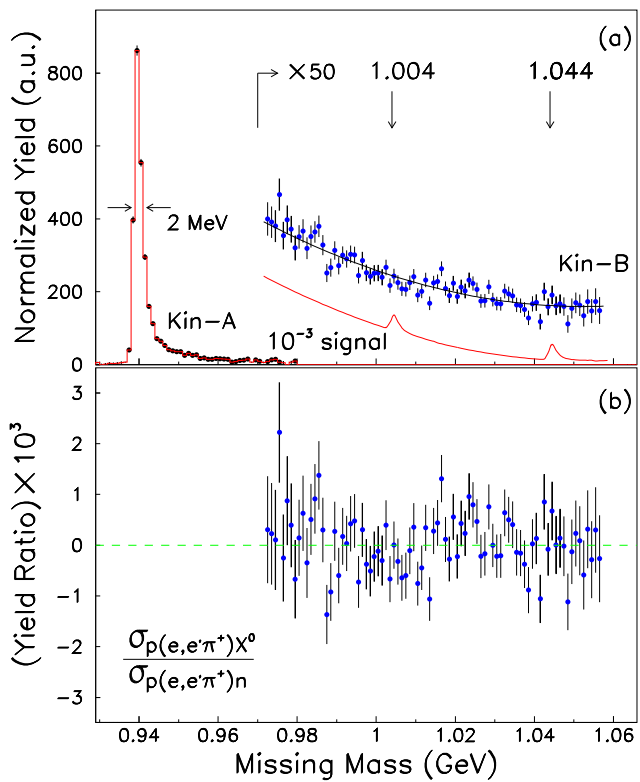,height=105mm,angle=0}
}
\caption{ Normalized $(e,e^{\prime}\pi^+)$ yields in arbitrary units are plotted in 1.0 MeV missing mass 
bins in (a).  
A third power polynomial fits the data, the fit residues 
 divided by the $p(e,e^{\prime} \pi^+)n$ peak-height are plotted in (b). Error bars are statistical only. 
 } 
\label{fig:mmiss} 
\end{figure}

The normalized yield, as a function of missing mass,  
is plotted in Fig.~\ref{fig:mmiss}(a) in 1.0 MeV bins. 
The missing mass resolution is 2.0 MeV,  as demonstrated in
Kinematics-A, and is mainly due to the energy loss in
the target material. Due to the large size of the target cell, the incident electron, 
the scattered electron, and 
the outgoing pion passed through averaged material thickness of 0.5, 0.6, and 0.4 g/cm$^2$, 
respectively.
 The yield of Kinematics-B, which is the Bethe-Heitler radiation tail of the 
$p(e,e^{\prime} \pi^+)n$ reaction, has been amplified fifty times in Fig.~\ref{fig:mmiss}(a).
A third-order polynomial was fit to the Kinematics-B yield with a reduced $\chi^2$ of 1.2 for 
 90 data points.
The signature of an $X^0$ resonance would be an excess of yield above the  smooth shape of the
 radiation tail. A line shape corresponding to such a signal with the 
strength of $K_s=1.0 \times 10^{-3}$ at 1.004 and 1.044 GeV 
is illustrated in Fig.~\ref{fig:mmiss}(a) by the curve shifted from the data.
The deviations of the data from the polynomial fit are divided by the $p(e,e^{\prime} \pi^+)n$ 
peak-height and plotted in Fig.~\ref{fig:mmiss}(b).
To the level of $K_s \approx 1.0 \times 10^{-3}$,  no resonance signal can be identified in the 
mass region of $0.97<M_{X^0}<1.06$ GeV.   Fitting these fluctuations with a Gaussian of 
FWHM $ = 2.0$ MeV, the experimental resolution, leads to $K_s=(-0.35 \pm 0.35) \times 10^{-3}$ at 1.004 GeV and
$K_s=(0.34 \pm 0.42) \times 10^{-3}$ at 1.044 GeV.
For the reported state at 0.987 GeV of Ref.~\cite{filkov},
we found $K_s=(-0.94 \pm 0.44) \times 10^{-3}$. 
The state at 1.094 GeV was not covered in this measurement.

The null result of this experiment does not necessarily 
contradict the claim of Ref.\cite{tati}.
First, the hadronic reaction  
could have allowed more exotic channels for the production of $X^0$, 
for example, through a reaction
with a dibaryon type intermediate state.  
Second, due to the limited beam time and the unfavorable target geometry, the sensitivity of this
measurement was not much improved compared to Ref.\cite{tati}.  
However, this measurement clearly demonstrated the potential of 
high resolution missing mass searches in coincidence  experiments.   
Further improvements of the experimental apparatus underway are expected to reduce 
the timing resolution to 0.5 ns and the  missing mass resolution to 0.5 MeV. 
A dedicated search\cite{prop} can set a tighter limit of $K_s \leq 1.0 \times 10^{-4}$. 
With the addition of neutron detectors or photon detectors 
to identify the radiative decay products of $X^0$, experimental sensitivity 
can be enhanced further by at least one order of magnitude.  
We point out that very weakly
excited states can only be observed when searched for seriously in dedicated experiments.
 This usually requires a great deal of care, effort and accelerator time, and most of all,
 very high resolution detectors.  Detailed attention must be paid to  
 backgrounds and instrumental effects which
 can lead to false structures. With the new generation of high resolution 
spectrometers and the high intensity CW electron 
beam of Jefferson Lab, a carefully planned experiment could set a much tighter limit on or even
discover narrow baryon structures which might not have been visible in earlier lower resolution experiments.

In conclusion, a high resolution missing mass search in the $p(e,e^{\prime}\pi^+)X^0$ reaction 
yielded no significant signal in the mass region of $0.97<M_{X^0} <1.06$ GeV.
The yield ratio of $p(e,e^{\prime}\pi^+)X^0/p(e,e^{\prime}\pi^+)n$ 
was determined to be $(-0.35 \pm 0.35) \times 10^{-3}$ at 1.004 GeV and
$(0.34 \pm 0.42) \times 10^{-3}$ at 1.044 GeV, consistent with a null signal.

We thank Drs.~E.~L.~Lomon, B.~Norum, B.~Tatischeff and W.~Turchinetz for many discussions.
We also thank the Jefferson Lab Hall A collaboration, Hall A technical staff and the Jefferson Lab Accelerator Division 
for their support of this experiment. This work was supported by the U.S. Department of Energy 
(contract DE-FG02-99ER41065, FIU)
and the National Science Foundation (contract PHY00-98642, Rutgers).
Southeastern Universities Research Association
manages Thomas Jefferson National Accelerator Facility
under DOE contract DE-AC05-84ER40150.

\end{document}